\documentclass[12pt]{iopart}

\newcommand{\be}{\begin{equation}}
\newcommand{\ee}{\end{equation}}
\newcommand{\bea}{\begin{eqnarray}}
\newcommand{\eea}{\end{eqnarray}}
\newcommand{\nn}{\nonumber}

\newcommand{\slsh}[1]{{\not \! #1}}
\newcommand{\TR}[1]{{{\rm Tr}\left[ #1 \right]}}

\begin{document}
\begin{flushright}
UMSNH-IFM-F-2008-26 
\end{flushright}

\title[Massive Dirac fermions and the zero field QHE]{Massive Dirac fermions and the zero field quantum Hall effect}
\author{Alfredo Raya$^{1}$ and Edward D Reyes$^{2}$}
\address{$^1$~Instituto de F\'{\i}sica y Matem\'aticas, Universidad Michoacana de San Nicol\'as de Hidalgo. Apartado Postal 2-82, C.P. 58040, Morelia, Michoac\'an, M\'exico.}
\address{$^2$~Instituto de Ciencias Nucleares, Universidad  Nacional Aut\'onoma de M\'exico. Circuito Exterior s/n,  Ciudad Universitaria, C.P. 04510, D.F., M\'exico.}
\ead{raya@ifm.umich.mx}

\begin{abstract}
Through an explicit calculation for a Lagrangian in quantum electrodynamics in (2+1)-space--time dimensions (QED$_3$), making use of the relativistic Kubo formula, we demonstrate that the filling factor accompanying the quantized electrical conductivity for massive Dirac fermions of a single species in two spatial dimensions is a half (in natural units) when time reversal and parity symmetries of the Lagrangian are explicitly broken by the fermion mass term. We then discuss the most general form of the QED$_3$ Lagrangian, both for irreducible and reducible representations of the Dirac matrices in the plane, with emphasis on the appearance of a Chern-Simons term. We also identify the value of the filling factor with a zero field quantum Hall effect (QHE).
\end{abstract}

\submitto{\JPA}
\maketitle

\section{Motivation}
One of the most remarkable phenomena in condensed matter physics discovered in the last decades of the past century is, indeed, the quantum Hall effect (QHE), a striking manifestation of quantum mechanics at the macroscopic level. The experimental observation of QHE is the vanishingly small diagonal electrical conductivity $\sigma_{xx}\to 0$ of a bidimensional non-relativistic electron gas subject to a strong perpendicular magnetic field, while the off-diagonal conductivity is quantized according to 
\be
\sigma_{xy}= - \nu\frac{e^2}{2\pi}\label{QHE}
\ee
in natural units ($\hbar=c=1$), where  $e$ is the electron charge and $\nu$ is the so-called filling factor. This is defined as the ratio of the density of electrons in the sample to the magnetic field strength, and can be a small integer, dubbed as Integer QHE (IQHE)~\cite{vonKiltzing}, or a fraction with odd denominator, the so called Fractional QHE (FQHE)~\cite{Tsui}. In view of the lack of relativistic corrections~\cite{MacDonald,Swamy1} to the quantization rule~(\ref{QHE}), a description of the QHE in terms of a full fledged relativistic quantum field theory is highly desirable. Quantum electrodynamics in (2+1)-space--time dimensions, QED$_3$, provides a natural framework for this purpose. This theory has useful applications, for example, in high-T$_c$ superconductivity~\cite{supercon} and in the field of dynamical chiral symmetry breaking, where  it provides an attractive battleground for lattice and continuum studies~\cite{dcsb}. More recently, QED$_3$ has been made use of in the description of the unconventional QHE in graphene~\cite{graphene}.  This material consists of an isolated single atomic layer of graphite, an ideal realization of a bidimensional system, which exhibits an unusual half-integer QHE in which massless ``relativistic''  carriers participate in the effect. A word of caution is at hand, since the term ``relativistic'' in the effective description of condensed matter systems of this kind has nothing to do with the familiar Lorentz symmetry in (2+1)-dimensions of the high energy physics (HEP) realm. Such a symmetry is not present in systems like graphene simply because the corresponding Dirac Lagrangian contains the Fermi velocity, $v_F$, instead of the velocity of light, $c$. This explains the use of the quotation marks. Charge carriers in graphene are ``relativistic'' in the sense that their energy-momentum dispersion relation is linear as opposed to the standard quadratic dispersion relation of non-relativistic systems.  In this paper, however, we give an idealized description of the QHE in terms of QED$_3$, which, of course, afterwards would have to be adapted to the real physical systems with their observed continuous and discrete symmetries.

An ideal tool for the computation of the transverse conductivity, and thus of the corresponding filling factor, is the Kubo formula~\cite{kubo}. Its field theoretical analog in the non-relativistic case was introduced in Ref.~\cite{Ishi} based upon gauge invariance  in terms of the Ward-Green-Takahashi identities~\cite{WGTI}. The filling factor is found through
\bea
\hspace{-2mm}
\nu &=& \frac{1}{24\pi^2} \int d^3p\  \epsilon_{\mu\rho\lambda} 
{\rm Tr}\left[
\partial^\mu S^{-1}(p) \ S(p) \ \partial^\rho S^{-1}(p) \ S(p)\ \partial^\lambda S^{-1}(p) \ S(p)\right] ,\label{filling}
\eea
where $S(p)$ is the electron propagator, $\epsilon_{\mu\rho\lambda}$ the Levi-Civita symbol and $\partial_\mu=\partial/\partial p^\mu$. The relativistic version of this expression preserves its functional form~\cite{Swamy1}, thus allowing a microscopic description of the QHE in terms of relativistic Dirac fermions. Pursuing this aim, Acharya and Swamy (hereafter referred to as AS) established that QED$_3$ naturally leads to the IQHE~\cite{Swamy1}. They considered non-interacting electrons, i.e., particles whose (inverse) propagator is simply given by
\be
S^{-1}(p)=\slsh{p}-m\;,\label{irrprop1}
\ee
where $m$ (assumed to be a positive quantity) is the effective mass of the electrons and they choose the following irreducible representation of the $\gamma$-matrices in terms of the Pauli matrices~:
\be
\gamma^0=\sigma^3, \qquad \gamma^1=i\sigma^1, \qquad \gamma^2=i\sigma^2, \label{rep1}
\ee
which verify $
\gamma^\mu \gamma^\nu = g^{\mu\nu}-i\epsilon^{\mu\nu\lambda}\gamma_\lambda$ with $\gamma_\mu = g_{\mu\nu} \gamma^\nu$.
This model has useful applications in the description of 1-dimensional polyacetylene~\cite{polyac}. In its full quantum field theoretical form, it exhibits the so-called ``parity anomaly''~\cite{Jackiw}, which has a condensed matter realization in a model which describes QHE without Landau levels~\cite{Hald}.
Inserting the propagator~(\ref{irrprop1}) into eq.~(\ref{filling}), with the aid of the properties of the Dirac matrices, 
\bea
\TR{\gamma^\mu}&=&0\nn\\
\TR{\gamma^\mu\gamma^\nu}&=&2g^{\mu\nu}\nn\\
\TR{\gamma^\mu\gamma^\nu\gamma^\lambda}&=&-2i \epsilon^{\mu\nu\lambda}\nn\\
\TR{\gamma^\mu\gamma^\nu\gamma^\lambda\gamma^\sigma}&=&2\left(g^{\mu\nu}g^{\lambda\sigma}+g^{\mu\sigma}g^{\nu\lambda}-g^{\mu\lambda}g^{\nu\sigma} \right)\nn\\
\TR{\gamma^\mu\gamma^\nu\gamma^\lambda\gamma^\sigma\gamma^\rho}&=&-2i\left(g^{\mu\nu}\epsilon^{\lambda\sigma\rho}+g^{\lambda\sigma}\epsilon^{\mu\nu\rho}+g^{\sigma\rho}\epsilon^{\mu\nu\lambda} -g^{\lambda\rho}\epsilon^{\mu\nu\sigma}\right)\;,
\label{traceirr}
\eea
AS find 
\bea
\nu &=&\frac{1}{24\pi^2}
\int d^3p \ \epsilon_{\mu\rho\lambda} \TR{\gamma^\mu \left( \frac{\slsh{p}+m}{p^2-m^2}\right)
\gamma^\rho \left( \frac{\slsh{p}+m}{p^2-m^2}\right)
\gamma^\lambda \left( \frac{\slsh{p}+m}{p^2-m^2}\right)}\nn\\
&=&\frac{1}{24\pi^2}
\int \frac{d^3p}{(p^2-m^2)^3} \ \epsilon_{\mu\rho\lambda} {\rm Tr}\left[\gamma^\mu\slsh{p}\gamma^\rho\slsh{p}\gamma^\lambda\slsh{p} +m^3\gamma^\mu\gamma^\rho\gamma^\lambda\right. \nn\\
&&+m\left(\gamma^\mu\slsh{p}\gamma^\rho\gamma^\lambda\slsh{p} + \gamma^\mu\gamma^\rho\slsh{p}\gamma^\lambda\slsh{p}+ \gamma^\mu\slsh{p}\gamma^\rho\slsh{p}\gamma^\lambda\right) \nn\\
&&\left.+m^2\left( \gamma^\mu\gamma^\rho\gamma^\lambda\slsh{p}+ \gamma^\mu\slsh{p}\gamma^\rho\gamma^\lambda+\gamma^\mu\gamma^\rho\slsh{p}\gamma^\lambda\right)\right]\nn\\
&=&\frac{1}{24\pi^2}
\int d^3p \frac{12im(p^2-m^2)}{(p^2-m^2)^3}\nn\\
&=& \frac{im}{2\pi^2} \int d^3p \frac{1}{(p^2-m^2)^2}=-\frac{1}{2}\;,\label{fill}
\eea
where the result in the last line comes after a Wick rotation to Euclidean space and a standard integration. Perhaps because this value of the filling factor does not correspond to the IQHE, nor to the FQHE, AS tried to give sense to their findings making three crucial statements:
\begin{enumerate}
\item Because electrons possess 2 spin states, a factor of 2 should be put by hand in order to obtain the IQHE.

\item The trace in eq.~(\ref{filling}) would have vanished for the usual QED$_4$ matrices.

\item The Chern-Simons term plays no role whatsoever.
\end{enumerate}
We find these statements misleading and, as such, these cast doubts on the results of works based on them~\cite{Swamy2,kurash}. Although we agree with AS in that the presence of the electron mass is crucial for QED$_3$ exhibiting ordinary QHE (and not the unconventional QHE found in graphene), we shall argue that these three statements cannot be correct. Criticisms to the ``factor of 2'' have already appeared in the literature~\cite{jellal,leitner}. Yet, our approach, based upon the analysis of the discrete symmetries (charge conjugation ${\cal C}$, parity $\cal P$, which for us is the inversion of one spatial axis, and time reversal ${\cal T}$)  of the  QED$_3$ Lagrangian, offers a new understanding of  the problems of these mishaps.

We have organized this article as follows: In Sect. 2 we discuss why the result in eq.~(\ref{fill}) is in fact correct, by analyzing the Dirac particle spectrum when irreducible representations of $\gamma-$matrices are chosen~\cite{Chuy,Shimizu}.  Section 3 is devoted to the calculation of the filling factor $\nu$ making use of the reducible representation of the usual $\gamma$-matrices of  QED$_4$, considering all types of fermion masses which can be present in (2+1)-dimensions.  We discuss the most general form of the QED$_3$ Lagrangian, with emphasis in the discrete symmetries (${\cal C, P, T}$) transformation properties of mass terms, both for electrons and photons (Chern-Simons term) in Sect. 4, and discuss how these can radiatively induce each other. Finally, in Sect. 5 we identify the half filling factor state for a single electron species as a zero magnetic field QHE  and then summarize our findings.

\section{Irreducible Dirac Fermions}

We start by noticing that the fermion propagator in eq.~(\ref{irrprop1}) can be read off from the (2+1)-analog of the usual free Dirac Lagrangian in 4D~:
\be
{\cal L}= \bar\psi(i\slsh{\partial}-m)\psi\;,\label{inhlag}
\ee
with the $\gamma$-matrices given in eq.(\ref{rep1}).  The spectrum of solutions of the resulting Dirac equation, which completes the corresponding Hilbert space, in the sense that the completeness relations $\sum u\bar{u}=\slsh{p}+m$ and $\sum v\bar{v}=\slsh{p}-m$ are fulfilled, is
\bea
\psi^P(x)&=& \left(\begin{array}{c} 
1 \\ \frac{p_x-ip_y}{E+m}
\end{array} \right) e^{-ix\cdot p} \equiv u(p)e^{-ix\cdot p} \nn\\
\psi^N(x)&=& \left(\begin{array}{c} 
 \frac{p_x+ip_y}{E+m} \\ 1
\end{array} \right) e^{ix\cdot p}\ \equiv v(p)e^{ix\cdot p}\;,
\eea
and consists of only a positive energy solution (particle spinor with spin up), and a negative energy one (antiparticle spinor with spin down)~\cite{Chuy}. Yet, these solutions fail to incorporate various symmetries of the ordinary Dirac spectrum in 4D. For example, the solutions, and correspondingly the Lagrangian in its \emph{inherited} form~(\ref{inhlag}), are not invariant under a Parity transformation ${\cal P}$ --which for consistency with Lorentz symmetry corresponds to the inversion of one spatial axis--,  nor under a time reversal transformation ${\cal T}$. Furthermore, only one out of the two spin states of the physical electrons is present in (2+1)-dimensions if we consider  an irreducible representation for the Dirac matrices. This fact makes it clear that the result in eq.~(\ref{fill}) is, in fact, correct and there is no need to put by hand the factor of 2 advocated in~\cite{Jackiw} and made use of in~\cite{Swamy1,Swamy2,kurash}. One might argue that since in condensed matter physics spin plays the role of flavor in HEP, one should in fact put the spin factor of 2 by hand. However, one cannot simply push this argument to the idealized Lagrangian of QED$_3$ we are considering here.  We shall postpone the discussion of the half filling state, and content ourselves at this stage with the filling factor vanishing or not. 

The two spin states and symmetry features of the familiar spectrum of solutions to the Dirac equation in 4D can be recovered owing to the fact that in (2+1)-dimensions there exists a second irreducible representation of the Dirac matrices in terms of the Pauli matrices, given by
\be
\gamma^0=\sigma^3, \qquad \gamma^1=i\sigma^1, \qquad \gamma^2=-i\sigma^2, \label{rep2}
\ee
with the property $\gamma^\mu \gamma^\nu=g^{\mu\nu}+i\epsilon^{\mu\nu\lambda}\gamma_\lambda$. 
Solutions of the Dirac equation in this representation are 
\bea
 \psi^P(x)&=& \left(\begin{array}{c} 
 \frac{p_x+ip_y}{E+m} \\ 1
\end{array} \right) e^{-ix\cdot p} \equiv u(p)e^{-ix\cdot p} \nn\\
\psi^N(x)&=& \left(\begin{array}{c} 
1\\ \frac{p_x-ip_y}{E+m}
\end{array} \right) e^{ix\cdot p}\ \equiv v(p)e^{ix\cdot p}
\;,
\eea
and correspond to particle spinor with spin down and antiparticle with spin up~\cite{Chuy}. These solutions fulfill the completeness relations $\sum u \bar{u}=\slsh{p}+m$ and $\sum v \bar{v}=\slsh{p}-m$, but present also the lack of spin states and symmetry properties of the familiar 4D solutions. Nevertheless, taking into account solutions for both representations, (\ref{rep1}) and (\ref{rep2}), attaching to them labels $A$ and $B$, respectively, we recover the features of the ordinary Dirac spectrum, namely, two spin states for the electrons and their corresponding Lorentz conjugated positron states. The two ``irreducible'' fermion fields can be cast into the following \emph{extended} form of the free Dirac Lagrangian~\cite{Chuy,Shimizu}~\footnote{Notice that only one irreducible representation of the Dirac matrices, say (\ref{rep1}) is used.}:
\be
{\cal L}= \bar\psi_A(i\slsh{\partial}-m)\psi_A+\bar\psi_B(i\slsh{\partial}+m)\psi_B\;.\label{extlag}
\ee
As we noticed before, neither under $\cal P$ nor under $\cal T$, the fields $\psi_A$ and $\psi_B$ transform onto themselves. In fact, under $\cal C$, $\cal P$ and $\cal T$ transformations, these fields transform as
\bea
(\psi_A)^{\cal C} = \gamma^2e^{i\eta_1} (\bar{\psi}_A)^T && \qquad (\psi_B)^{\cal C} = \gamma^2e^{i\eta_2} (\bar{\psi}_B)^T\nn\\
(\psi_A)^{\cal P} = -i \gamma^1e^{i\phi_1} (\psi_B) && \qquad (\psi_B)^{\cal P} = -i \gamma^1e^{i\phi_2} (\psi_A)\nn\\
(\psi_A)^{\cal T} = i \gamma^3e^{i\varphi_1} (\bar{\psi}_B)^T && \qquad (\psi_B)^{\cal P} = -i \gamma^3e^{i\varphi_2} (\bar{\psi}_A)^T,
\eea
where $\eta_i$, $\phi_i$ and $\varphi_i$, $i=1,2$ are constant phases. This shows that the extended Lagrangian (\ref{extlag}) is $\cal CPT$ invariant~\cite{Shimizu}. The model~(\ref{extlag}) has recently been considered to study  the formation of $\bar\psi \psi$-condensates in the presence of magnetic fields even in the absence of fermion masses~\cite{ABR}. For this Lagrangian, the fermion propagators are of the form~(\ref{irrprop1}), but now $m$ alternates in sign between $\psi_A$ and $\psi_B$. 
Thus, since there are two fermion species in the La\-gran\-gian~(\ref{extlag}), the filling factor is~\cite{hatsugai}
\be
\nu= -\frac{1}{2} \sum_\alpha {\rm sgn}(m_\alpha) \label{species}
\ee
where $m_\alpha$ is the mass of fermion species $\alpha$, and thus vanishes in this case. This result is understandable because $\sigma_{xy}$ is a $\cal P$ and $\cal T$ violating quantity, whereas the Lagrangian in eq.~(\ref{extlag}), from which we are deriving it, is not. Thus $\sigma_{xy}$, or equivalently $\nu$, can only be zero. This will be further clarified  in Sect. 4.

The presence of two irreducible fermion fields in~(\ref{extlag}) naturally suggest that these can be merged into one reducible four-component spinor and hence we can make use of the ordinary $4\times 4$ Dirac matrices of QED$_4$. Nevertheless, care should be taken since, in (2+1)-dimensions, further mass terms, besides the ordinary $m\bar\psi\psi$, can arise. Such an issue is discussed below.

\section{Reducible Dirac Fermions}

If ordinary $4\times 4$ Dirac matrices are made use of, only three of them are required to describe the dynamics of electrons in (2+1)-dimensions, for example $\{\gamma^0,\ \gamma^1, \ \gamma^2\}$, which can be represented as
\be
\gamma^0=\left(\begin{array}{cc} \sigma^3 & \phantom{m}0 \\ 0 & -\sigma^3\end{array} \right), \quad
\gamma^1=\left(\begin{array}{cc} i\sigma^1 & \phantom{m}0 \\ 0 & -i\sigma^1\end{array} \right), \quad
\gamma^2=\left(\begin{array}{cc} i\sigma^2 & \phantom{m}0 \\ 0 & -i\sigma^2\end{array} \right).
\ee
In that case, we have two other $\gamma$ matrices which commute with all the three matrices above, in such a fashion that the corresponding massless Dirac Lagrangian is invariant under the chiral-like transformations 
$\psi \to e^{i\alpha \gamma^3}\psi \;,$ and $ \psi \to e^{i\alpha \gamma^5}\psi\;, $
that is, it is invariant under a global $U(2)$ symmetry with generators $1$, $\gamma^3$, $\gamma^5$ and $[\gamma^3, \gamma^5]$. Here 
\be 
\gamma^3=i\left(\begin{array}{cc} 0 & I \\ I & 0 \end{array} \right), \qquad
\gamma^5=i\gamma^0\gamma^1\gamma^2\gamma^3=\left(\begin{array}{cc} \phantom{-}0 & i \\ -i & 0 \end{array} \right)\;,
\ee
$I$ being the $2\times 2$ unit matrix. This symmetry is broken by an ordinary mass term $m_e\bar\psi\psi$. But there exists another mass term (sometimes referred to as Haldane mass term) which is invariant under the ``chiral'' transformations
\be
m_o \bar\psi \frac{i}{2} [\gamma^3, \gamma^5] \psi \equiv m_o \bar\psi (i\tau) \psi.\label{taumass}
\ee
If we write the 4-spinor as
\be
\psi=\left( \begin{array}{c}\psi_1 \\ \psi_2 \end{array}\right)\;,
\ee
we observe that under $\cal P$ and $\cal T$, the upper and lower components of this spinor transform, up to a phase, as~\cite{JT}
\bea
\left(\psi_1(t,x,y)\right)^{\cal P}\to \sigma^1\psi_2(t,-x,y)&& \qquad \left(\psi_2(t,x,y)\right)^{\cal P}\to \sigma^1\psi_1(t,-x,y)\nn\\
\left(\psi_1(t,x,y)\right)^{\cal T}\to \sigma^2\psi_2(-t,x,y)&& \qquad \left(\psi_2(t,x,y)\right)^{\cal T}\to \sigma^2\psi_1(-t,x,y).
\eea
Thus, the term $m_e\bar\psi\psi$ is even under each of these transformations, but $m_o \bar\psi (i\tau) \psi$ is not, although it is $\cal PT$ and thus $\cal CPT$ symmetric. Here the evenness and oddness of the mass terms under $\cal P$ and $\cal T$ justify the use of the subscripts ``$e$'' and ``$o$''. The corresponding Euclidean space free \emph{reducible} Dirac Lagrangian in this case has the form
\be
{\cal L}= \bar\psi(i\slsh{\partial} -m_e-m_o\tau)\psi\;,\label{redlag}
\ee
where the Euclidean Dirac matrices are chosen as
\be
\gamma^0=\left(\begin{array}{cc} -i\sigma^3 & 0 \\ 0 & i\sigma^3\end{array} \right), \quad
\gamma^1=\left(\begin{array}{cc} i\sigma^1 & \phantom{m}0 \\ 0 & -i\sigma^1\end{array} \right), \quad
\gamma^2=\left(\begin{array}{cc} i\sigma^2 & \phantom{m}0 \\ 0 & -i\sigma^2\end{array} \right),
\ee
such that
\be 
\gamma^3=\left(\begin{array}{cc} 0 & I \\ I & 0 \end{array} \right), \quad
\gamma^5=\left(\begin{array}{cc} 0 & -I \\ I & \phantom{-}0 \end{array} \right),\quad 
\tau= \left(\begin{array}{cc} I & 0 \\ 0 & -I \end{array} \right).
\ee
There are many planar condensed matter models in which the low energy sector can be written as this effective form of QED$_3$, for which the physical origin of the masses depends on the underlying system~\cite{Sharapov}. For example $d$-wave cuprate superconductors~\cite{supercon}, $d$-density-wave states~\cite{ddw}, and layered graphite~\cite{lg}, including graphene in the massless version~\cite{graphene}. The reader should bear in mind that the discrete ${\cal P}$ and ${\cal T}$ symmetries discussed above do not have direct relation to corresponding symmetries in two-dimensional condensed matter systems. For example, there is no symmetry with respect to a reflection of one spatial coordinate in graphene. Instead the space-inversion symmetry contains reflection of signs of two spatial coordinates and the exchange of the $A$ and $B$ atoms of the honeycomb lattice and ${\mathbf K}_\pm$ points in the Brillouin zone. On the other hand, the time reversal operation, which flips the spin signs, interchanges ${\mathbf K}_\pm$, but not sublattices.

In order to perform a calculation of the filling factor, it is convenient to introduce the chiral-like projectors
\be
\chi_\pm = \frac{1}{2}(1\pm \tau)\;,
\ee
which verify~\cite{Kondo}
$\chi_\pm^2=\chi_\pm$, $\chi_+\chi_-=0$, $\chi_++\chi_-=I$. The ``right handed'' $\psi_+$ and ``left handed'' $\psi_-$ fermion fields in this case are given by $\psi_\pm=\chi_\pm \psi$. The $\chi_\pm$ project the upper and lower two component spinors (fermion species) out of the four-component $\psi$. The chiral-like decomposition of the free fermion propagator is 
\bea
S(p) &=& - \left(\frac{\slsh{p}+m_+}{p^2+m_+^2} \chi_+ + \frac{\slsh{p}+m_-}{p^2+m_-^2} \chi_- \right)\nn\\
& \equiv &- \left(S_+(p) \chi_+ +S_-(p) \chi_-\right)\;,\label{redupropchi}
\eea
where $m_\pm = m_e \pm m_o$. Hence, the right- and left-handed projections of the filling factor can be obtained from eq.~(\ref{filling}) as
\bea
\nu_\pm &=& \frac{1}{24\pi^2} \int d^3p\  \epsilon_{\mu\nu\lambda} 
{\rm Tr}\left[
\partial^\mu S^{-1}_\pm  S_\pm   
\partial^\rho S^{-1}_\pm S_\pm \partial^\lambda S^{-1}_\pm  S_\pm \chi_\pm \right] \;,
\label{fillingred}
\eea
where we have omitted the dependence of the propagators on the fermion momentum $p$ to avoid cumbersome notation. With the aid of the relations
\bea
\TR{\chi_\pm}&=&2\;,\nn\\
\TR{\gamma^\mu\chi_\pm}&=&0\nn\\
\TR{\gamma^\mu\gamma^\nu\chi_\pm}&=&-2\delta^{\mu\nu}\nn\\
\TR{\gamma^\mu\gamma^\nu\gamma^\alpha\chi_\pm}&=&\mp2\epsilon^{\mu\nu\alpha}\nn\\
\TR{\gamma^\mu\gamma^\nu\gamma^\alpha\gamma^\beta\chi_\pm}&=&2\left(\delta^{\mu\nu}\delta^{\alpha\beta}+\delta^{\mu\beta}\delta^{\nu\alpha}-\delta^{\mu\alpha}\delta^{\nu\beta} \right)\;,
\eea
the calculation of (\ref{fillingred}) becomes very similar to that in eq.~(\ref{fill})~:
\bea
\nu_\pm &=&\frac{\epsilon_{\mu\rho\lambda}}{24\pi^2}
\int d^3p \  \TR{\gamma^\mu \left( \frac{\slsh{p}+m_\pm}{p^2+m^2_\pm}\right)
\gamma^\rho \left( \frac{\slsh{p}+m_\pm}{p^2+m^2_\pm}\right)
\gamma^\lambda \left( \frac{\slsh{p}+m_\pm}{p^2+m^2_\pm}\right)\chi_\pm}\nn\\
&=&\frac{\epsilon_{\mu\rho\lambda}}{24\pi^2}
\int \frac{d^3p}{(p^2+m_\pm^2)^3} \  {\rm Tr}\left[\gamma^\mu\slsh{p}\gamma^\rho\slsh{p}\gamma^\lambda\slsh{p}\chi_\pm +m_\pm^3\gamma^\mu\gamma^\rho\gamma^\lambda\chi_\pm\right. \nn\\
&&+m_\pm\left(\gamma^\mu\slsh{p}\gamma^\rho\gamma^\lambda\slsh{p} + \gamma^\mu\gamma^\rho\slsh{p}\gamma^\lambda\slsh{p}+ \gamma^\mu\slsh{p}\gamma^\rho\slsh{p}\gamma^\lambda\right)\chi_\pm \nn\\
&&\left.+m_\pm^2\left( \gamma^\mu\gamma^\rho\gamma^\lambda\slsh{p}+ \gamma^\mu\slsh{p}\gamma^\rho\gamma^\lambda+\gamma^\mu\gamma^\rho\slsh{p}\gamma^\lambda\right)\chi_\pm\right]\nn\\
&=& \mp \frac{1}{24\pi^2} \int d^3p \frac{12m_\pm (p^2+m_\pm^2)}{(p^2+m_\pm^2)^3}\nn\\
&=&\mp \frac{m_\pm}{2\pi^2}\int d^3p \frac{1}{(p^2+m_\pm^2)^2}=\mp \frac{1}{2}{\rm sgn}(m_\pm)\;. \label{nured}
\eea
This proves the usefulness of the chiral-like projectors. Notice that these projectors 
only account for the same degrees of freedom in a different basis.

From the above expressions, we obtain that 
\be
\nu = \nu_{+} + \nu_{-} = - \frac{1}{2}{\rm sgn}(m_{+}) +\frac{1}{2}{\rm sgn}(m_{-}).
\ee
This gives $\nu=0$ if $m_{o}=0$, and $\nu= -1$ if $m_{e}=0$, -1/2 coming from each fermion species.
Two comments are in order at this point: First, we have seen that even with the use of ordinary $\gamma^\mu$-matrices, the trace in eq.~(\ref{fillingred}) yields a nonvanishing result, contrary to the claims of AS, \cite{Swamy1}. This is so because of the presence of the mass term~(\ref{taumass}),  which violates $\cal P$ and $\cal T$ just like the ordinary fermion mass term in the irreducible Dirac Lagrangian~(\ref{inhlag}). Not surprisingly, this result comes about since the physical content of eqs.~(\ref{filling}) and~(\ref{fillingred}) is the same, regardless of the representation of the $\gamma^\mu$-matrices. Second, since in the reducible representation of the Dirac matrices two different species of electron fields are taken into account, each contributing with -1/2 to the filling factor, QED$_3$ with its full-fledged relativistic symmetry, naturally leads to the IQHE. The reader should be careful in the interpretation of  this result. As we previously pointed out, in condensed matter physics spin plays the role of flavor in HEP and two-dimensional Dirac fermions, combined into a four-dimensional Dirac spinor, describe gapless excitations near different nodal points of the Fermi surface, which in graphene, for example, belong to the ${\mathbf K}_\pm$ points in the extended Brillouin zone. In the absence of a magnetic field, one in fact has to take the spin factor 2 by hand. In the case we are considering here, such a factor comes from the flavor of fermions.

Notice that it is only the $\cal P$ and $\cal T$ odd mass term, both in the irreducible~(\ref{inhlag}) and reducible~(\ref{redlag}) Lagrangians which leads to a nonvanishing quantized conductivity, 
in agreement with the general properties required for a theory to exhibit QHE~\cite{Avron}. This fact is important because of the Coleman-Hill theorem~\cite{CH}, which states that such mass term radiatively gives rise to a Chern-Simons term only at the one-loop level. So, it is desirable that such term appears directly in the bare Lagrangian. We shall now discuss the most general Lagrangian of QED$_3$ under this observation and we argue why the third comment of AS  cannot be correct either.

\section{QED$_3$ Lagrangian} 

To start with the discussion of the QED$_3$ Lagrangian, let us recall that its 4D analog is of the form
\be
{\cal L}_{QED}= {\cal L}_{Dirac}+{\cal L}_{\gamma}+{\cal L}_{GF}+{\cal L}_{Int}\;, \label{lag4D}
\ee
where
\be 
{\cal L}_{\gamma}=-\frac{1}{4}F_{\mu\nu}F^{\mu\nu}\;,
\ee
 ${\cal L}_{GF}$ is the gauge fixing Lagrangian, whose form is irrelevant in our discussion, and ${\cal L}_{Int}$ is the interaction term which minimally couples matter fields to the photon field. This form, however, is not the most general  in (2+1)-dimensions. Firstly, as we have seen in previous sections, we can work with reducible and irreducible representations for the $\gamma^\mu$-matrices. If we insist on sticking to a single irreducible representation, like the one in eq.~(\ref{rep1}), the ordinary mass term $m\psi\bar\psi$ breaks both the $\cal P$ and $\cal T$ invariance of ${\cal L}_{Dirac}$. Furthermore, because of the Coleman-Hill theorem~\cite{CH}, this mass term (and any other odd mass term), would radiatively generate a Chern-Simons term (CST)
\be
{\cal L}_{CS} = \frac{\vartheta}{4} \epsilon_{\mu\rho\lambda}F^{\mu\rho}A^\lambda\label{TCS} 
\ee
into the Lagrangian of QED$_3$, which induces a gauge invariant topological mass for the photon. In order to see that, consider,  the Lagrangian (\ref{lag4D}) with ${\cal L}_{Dirac}$ given in~(\ref{inhlag}). The first radiative correction to the current-current correlation function, out of which one can also compute the transverse conductivity (see Appendix), and that in the HEP language is known as the vacuum polarization, is given by
\be
\Pi_{\mu\rho}(q)=  ie^2 \int \frac{d^3k}{(2\pi)^3} \TR{\gamma^\mu \frac{1}{\slsh{k}-m}\gamma^\rho \frac{1}{\slsh{k}+\slsh{q}-m} }\;.\label{vac}
\ee
One can see from the properties of the Dirac matrices~(\ref{traceirr}) that the tensor structure of the vacuum polarization is
\bea
\Pi_{\mu\rho}(q)&=&\Pi_{\mu\rho}^{e}(q) + \Pi_{\mu\rho}^o(q)\nn\\
&=&\left(g_{\mu\rho}-\frac{q_\mu q_\rho}{q^2} \right)\Pi_e(q^2)+im\epsilon_{\mu\rho\lambda}q^\lambda \Pi_o(q^2)\;,
\label{tensorpol}
\eea
where we have used once more the labels $e$ and $o$ for the even and odd parts of the vacuum polarization under $\cal P$ and $\cal T$ transformations and the factor $m$ is used for later convenience. The tensor structure of the second term, which emerges from the trace of three $\gamma$ matrices, induces a Chern-Simons interaction $\epsilon_{\mu\rho\lambda} F^{\mu\rho}A^\lambda$, which at the level of Lagrangians corresponds to~(\ref{TCS}). A similar argument follows for  ${\cal L}_{Dirac}$ of the form~(\ref{redlag}). As we shall see in the Appendix, the linear part of $\Pi_o(q^2)$ is related to the Hall conductivity and thus to the filling factor. With the $\cal P$ and $\cal T$ preserving Lagrangian~(\ref{extlag}), the vacuum polarization has only the even tensor structure, i.e., $\Pi_o(q^2)\equiv 0$, which readily implies $\sigma_{xy}=0$ and thus $\nu=0$. 

The CST also violates $\cal P$ and $\cal T$, but is $\cal PT$ and thus $\cal CPT$ invariant (see, for example,~\cite{khare}). The converse is also true, if we start from a Lagrangian of QED$_3$ which includes the CST but no $\cal P$ and $\cal T$ violating fermion mass term, the CST would radiatively generate it~\cite{Del}, in such a fashion that fermions acquire their usual relativistic energy spectrum, with a mass gap. Observe that in the case of the Dirac Lagrangian~(\ref{extlag}), no CST would be radiatively induced, because the two fermion fields appear with different signs for their mass terms. Thus, depending upon the choice of the representation for the Dirac matrices, we get three possible choices for the QED$_3$ Lagrangian~:
\begin{itemize}
\item \emph{Case I  ~:}  For the \emph{inherited} Dirac Lagrangian~(\ref{inhlag}), 
\bea
{\cal L}_{QED_3}^I
&=& \bar\psi\left(i\slsh{\partial}-m\right) \psi  +{\cal L}_{\gamma}+{\cal L}_{GF}+ {\cal L}_{Int}
+ \frac{\vartheta}{4} \epsilon_{\mu\nu\lambda} A^\mu F^{\nu\lambda} \;.\label{IQED3}
\eea
In this case, the mass term and the CST induce each other mutually. This is the most general form of the Lagrangian if the representation (\ref{rep1}) or (\ref{rep2}) is chosen.\\

\item \emph{Case II ~:}  For the \emph{extended} Dirac Lagrangian~(\ref{extlag}),
\bea
{\cal L}_{QED_3}^E
&=& {\bar\psi}_A\left(i\slsh{\partial}-m\right) \psi_A  
+ {\bar\psi}_B\left(i\slsh{\partial}+m\right) \psi_B
+{\cal L}_{\gamma}+{\cal L}_{GF}+ {\cal L}_{Int}^{A,B}
\label{EQED3}
\eea
In this case there is no CST, because the signs of the fermion mass terms cancel between them any contribution to~(\ref{TCS}).\\

\item \emph{Case III~:}  For the \emph{reducible} Dirac Lagrangian~(\ref{redlag}),
\bea
{\cal L}_{QED_3}^R
= \bar\psi\left(i\slsh{\partial}\!-\!m_e\!-\!m_o\tau \right) \psi \!+\!{\cal L}_{\gamma}
 \!+\!{\cal L}_{GF}\!+\! {\cal L}_{Int}
\!+\! \frac{\vartheta}{4} \epsilon_{\mu\nu\lambda} A^\mu  F^{\nu\lambda} .\label{RQED3}
\eea
Here $m_o$ and the CST induce each other  mutually.\\

\end{itemize}

Thus, as we have seen in previous sections, only those $\cal P$ and $\cal T$ violating Lagrangians, (\ref{IQED3}) and (\ref{RQED3}), give nonvanishing filling factors. Fermion masses which break $\cal P$ and $\cal T$  are responsible for such contributions, and these and the CST generate each other radiatively. Then,  the sole presence of $m$ in (\ref{IQED3}) or $m_o$ in (\ref{RQED3}) unveils the presence of the CST, a statement that contradicts the third comment of AS. Below we discuss the consequences of the symmetries of these Lagrangians and their connection to the nonvanishing filling factor.

\section{Summary and Discussion}

When and if the Lagrangian for electrons, restricted to live in a plane, is not invariant under the discrete symmetries of $\cal P$ and $\cal T$, a non-vanishing transverse conductivity develops and QHE emerges~\cite{Avron}. In this fashion, IQHE takes place as a result of the quantization conditions for individual electrons in a magnetic field. The number of filled Landau levels, the filling factor $\nu$, is related to the induced effective action for the gauge field, namely, the CST, in the vacuum of the electrons. If interactions between electrons play a role, bulk effects are manifest, setting the scene for the FQHE to take place. In this effect, because of interactions between electrons, Landau levels are filled only partially. This can be visualized as if electrons were subject to an effective magnetic field, the external plus a fictitious field arising from their many-body wavefunction statistical properties. In the particular case when Landau levels are half filled, such an effective magnetic field vanishes. Thus, there is a duality between a system of electrons in a magnetic field at $\nu=1/2$ and a system of non-interacting electrons which exhibits QHE at zero magnetic field~\cite{leitner,frohlich}. 
As such, this duality  can be exploited to obtain physical properties in a simpler manner.


For example,  it was experimentally de\-mons\-tra\-ted that in a usual quantum Hall system, the $\nu=1/2$ state is realized either in a wide single quantum well or a double quantum well~\cite{Eisenstein}.  Resonance of a surface sound wave with cyclotron orbits of charge carriers was also observed in this system, although perhaps the most striking observation was that this system develops a well defined Fermi surface~\cite{Willet}. The theoretical explanation for the formation of such a Fermi surface and other features observed in this state was proposed~\cite{HLR} precisely making use of this duality. Electrons in this case were considered non-relativistic in nature. The physical  picture  is that this system is an electrical dipole created by electrons and vortices~\cite{stormer}. 
But also models of massive relativistic Dirac fermions have been considered as a realization of this duality. The role of a Dirac mass term of the $\cal P$ and $\cal T$ violating type  (Haldane mass) was first studied on the lattice to explain a QHE without Landau levels~\cite{Hald}. Later,  such a mass  was shown to be important to explain the nature of transitions in the IQHE~\cite{Lud,hatsugai}, playing the role of the energy in the usual treatment of the IQHE. In all these cases, the filling factor associated to massive Dirac fermions of a single species was found to be a half, in agreement with our findings. There is no problem with that, because a single Dirac fermion is incapable of reproducing the behavior of a system with a finite number of degrees of freedom per volume. Lattice results suggest that there is always  an even number of particles participating in the effect, at least one of them being a massive spectator~\cite{hatsugai}. More recently, the $\nu=1/2$ result of the zero field QHE has been nicely explained in geometrical terms~\cite{leitner}, identifying the corresponding conductivity with a solid angle, clarifying the problems with the ``factor of 2'' of AS. 

The identification of the discrete symmetries of the QED$_3$ Lagrangian, related to the different representations which can be given to the Dirac matrices and the appearance of the CST, set the basis of a complete understanding of the system of non-interacting fermions which exhibits QHE without Landau levels. The duality between systems of interacting and non-interacting fermions, in the light of the emergence of ``relativistic'' condensed matter systems, brings winds of optimism to the solution of current puzzles, like the minimal conductivity in graphene~\cite{kat}.

In summary, we have undertaken an idealized microscopical description of the QHE in a full-fledged relativistic quantum field theory, QED$_3$, by means of the field theoretical version of the Kubo formula, eq.~(\ref{filling}), written in terms of the fermion propagator. We have observed that this theory naturally leads to a half filling QHE per fermion species present in the underlying Lagrangian, when $\cal P$ and $\cal T$ are explicitly broken by a fermion mass term. In the process, we have examined three critical statements made by AS in Ref.~\cite{Swamy1} from a Lagrangian point of view. We have found that for a single fermion species:
\begin{itemize}
\item There is no need to multiply  this result by 2.
\item The same result can be obtained considering the ordinary Dirac matrices, but considering the usual fermion mass term and a $\cal P$ and $\cal T$ violating mass term, eq.~(\ref{taumass}).
\item CST is implicit in the result.
\end{itemize}
Indeed,  if an irreducible representation of the $\gamma^\mu$ matrices is made use of, (i) the ordinary mass term $m\bar\psi\psi$ is odd under $\cal P$ and $\cal T$; and (ii) the ordinary spectrum of solutions of the Dirac equation, although complete --in the sense that completeness relations are fulfilled--, lacks two polarization states for the electron~\cite{Chuy}. One of the most significant features of the ordinary spectrum in 4D, namely, the existence of two spin states for electrons and positrons, can be recovered owing to the fact that there exists a second irreducible representation of the Dirac matrices which enforces us to consider two fermion species in an ``extended'' Lagrangian~\cite{Chuy,Shimizu}, eq.~(\ref{extlag}); however, this yields a vanishing filling factor. Also, merging two different fermion species, making use of a reducible representation for the Dirac matrices, the Kubo formula yields a half filling factor per species, provided a $\cal P$ and $\cal T$ violating mass term, eq.~(\ref{taumass}), is included. Thus, we confirm that in order for a theory to exhibit QHE, its underlying Lagrangian must be   $\cal T$ (and $\cal P$) violating, but $\cal CPT$ preserving~\cite{leitner,Avron,frohlich}. In our case, the fermion mass term, $m$ in (\ref{inhlag}) and $m_o$ in (\ref{redlag}), are responsible for this. In the case of graphene,  where massless Dirac fermions participate in the effect, the external magnetic field is responsible for breaking the discrete symmetries. The vanishing of the filling factor in graphene in the zero field limit has been shown by lattice simulations~\cite{Hald} and by continuum studies~\cite{GusShar}. 
Furthermore, a $\cal P$ and $\cal T$ violating fermion mass term radiatively generates a CST~\cite{CH} and vice versa~\cite{Del}, rendering the energy spectrum of massive Dirac fermions in its usual relativistic (gapped) form. Thus, the sole presence of such fermion mass,  implies presence of the CST in the underlying Lagrangian. The existence of the CST implies a Hall conductivity, in this case characterized by a filling factor $\nu=1/2$ per species. The non-integer nature of this $\nu$ suggest its origin as a bulk effect of a system of interacting fermions. Yet it was computed considering non-interacting electrons. This apparent inconsistency is solved by a duality argument: a system of interacting electrons of a single species in a magnetic field at half filling is visualized as a  bidimensional gas of electrons at zero magnetic field which exhibits QHE. This  state  is pictured as an electrical electron-vortex dipole~\cite{HLR}. 
 
In this fashion, the findings of AS, eq.~(\ref{fill}), which we have proved to be correct, describe a very interesting physical system.  Their misinterpretations, presumably stated to enforce their results to the IQHE, cannot be correct. Hence, the findings of papers based on AS reasoning,~\cite{Swamy2,kurash}, specially in the description of the FQHE, need further revision. A key observation for the description of QHE for systems of relativistic fermions is the non invariance of the mass terms under spatial and time reflections. This becomes even more relevant in systems like graphene, where a plethora of mass terms with a variety of space-time transformation properties can be considered~\cite{GusRev}. The complete identification of such mass terms in the dynamics of QHE in graphene is in progress.

\ack
We acknowledge Alejandro Ayala, Adnan Bashir, Robert Delbourgo, Mariana Kirchbach and Manuel Torres  for valuable discussions and careful reading of the manuscript. Support has been received from SNI, COECyT and CIC under projetcs CB0702162\_4 and 4.22 respectively.

\appendix
\section*{Appendix}
\setcounter{section}{1}
The Hall conductivity is defined as the linear part of the current-current correlation function~\cite{Swamy1}~:
\be
\sigma_{xy}=\frac{1}{3!} \epsilon_{\mu\rho\lambda} \left.\frac{\partial}{\partial q^\lambda} \Pi^{\mu\rho}(q^2)\right|_{q^2=0}\;.
\ee
From the tensor structure of the vacuum polarization~(\ref{tensorpol}), we see that the symmetric part of the vacuum polarization, $\Pi^{\mu\nu}_e$,  vanishes upon contraction with the Levi-Civita symbol. Thus
\bea
\sigma_{xy}&=&\frac{1}{3!} \epsilon_{\mu\rho\lambda} \left.\frac{\partial}{\partial q^\lambda}\left(im\epsilon^{\mu\rho\eta}q_\eta\Pi_o(q^2) \right)\right|_{q^2=0}\nn\\
&=& im\Pi_o(0)+\frac{im}{3!} q_\eta \left.\frac{\partial}{\partial q^\eta}\Pi_o(q^2)\right|_{q^2=0}\;.
\eea
Now, from the vacuum polarization~(\ref{vac}), we have that
\bea
\Pi_o(q^2)&=&\frac{-ie^2}{4\pi^3} \int d^3k \frac{1}{[k^2-m^2][(k+q)^2-m^2]}\nn\\
&=&\frac{-i e^2}{4\pi\sqrt{q^2}}\ln{\left( \frac{2m+\sqrt{q^2}}{2m-\sqrt{q^2}}\right)} \;.
\eea
From this expression, it is not difficult to see that 
\be
\Pi_o(0)=\frac{-i e^2}{4\pi m}\;, \qquad q^2 \left.\frac{\partial}{\partial q^2}\Pi_o(q^2)\right|_{q=0}=0\;,
\ee
and thus
\be
\sigma_{xy} = im\Pi_o(0) = \frac{e^2}{4\pi}\;.
\ee
Comparing with~(\ref{QHE}), we again obtain $\nu=-1/2$. Although we have used an irreducible representation of the Dirac matrices in this calculation, the same result per fermion species holds with a reducible representation, provided we consider the term $m_0 \bar\psi \tau \psi$. Furthermore, notice that $\Pi_o(0)$ is precisely the coefficient of the CST in~(\ref{TCS}), thus the non vanishing value of filling factor is related to the non vanishing of this coefficient, as we stated earlier.


\section*{References}

\begin{thebibliography}{99}
\bibitem{vonKiltzing} K. von Kiltzing, G. Dorda and M. Pepper, Phys. Rev. Lett. {\bf 45}  494 (1980).
%
\bibitem{Tsui} DC. Tsui, H.L. Stormer and A.C. Gossard, Phys. Rev. Lett. {\bf 48},  1559 (1982).
%
\bibitem{MacDonald} A.H. MacDonald, Phys. Rev. {\bf B28}, 2235 (1983);
 M. Martin Nieto and P.L. Taylor. Am. J. Phys {\bf 53}, 234  (1985).
%
\bibitem{Swamy1} R. Acharya and P. Narayana Swamy, Il Nouvo Cimento {\bf B107}, 351 (1992).
%
\bibitem{supercon} N. Dorey and N.E. Mavromatos, Nucl. Phys. {\bf B386}, 614 (1992); 
K. Farakos and N.E. Mavromatos, Mod. Phys. Lett. {\bf A13}, 1019 (1998); 
M. Franz and Z. Tesanovic, Phys. Rev. Lett. {\bf87}, 257003 (2001); 
O. Vafek, A. Melikyan, M. Franz and T. Tesanovic, Phys. Rev. {\bf B63}, 134509 (2001);
I.F. Herbut, Phys. Rev. {\bf B66}, 094504 (2002);
M. Franz, Z. Tesanovic and O. Vafek, Phys. Rev. {\bf B66}, 054535 (2002); 
M. Sutherland {\em et. al.}, Phys. Rev. Lett. {\bf 94},  147004 (2005).
%
\bibitem{dcsb} 
T. Appelquist, M.J. Bowick, D. Karabali and L.C.R. Wijewardhana, Phys. Rev. {\bf D33},  3704 (1986);
M.R. Pennington and D. Walsh, Phys. Lett. {\bf B253},  246 (1991); 
C.J. Burden and C.D. Roberts, Phys. Rev. {\bf D44},  540  (1991);
D.C. Curtis, M.R. Pennington and D. Walsh, Phys. Lett. {\bf B295},  313 (1992);
S.J. Hands, J.B. Kogut and C.G. Strouthos, Nucl. Phys. {\bf B645}, 321  (2002); 
 A. Bashir, A. Huet and A. Raya.  Phys.  Rev. {\bf D66},  025029  (2002);
C.G. Strouthos, Nucl. Phys. Proc. Suppl. {\bf 119},  974  (2003);
V.P. Gusynin and M. Reenders, Phys. Rev. {\bf D68},  025017  (2003); 
S.J. Hands, J.B. Kogut, L. Scorzato and C.G. Strouthos, Phys. Rev. {\bf B70}, 104501  (2004);
C.S. Fischer, R. Alkofer, T. Dahm and P. Maris, Phys. Rev. {\bf D70},  073007  (2004);
Y. Hoshino. JHEP {\bf 0409},  048  (2004);
A. Bashir and A. Raya, Nucl. Phys. {\bf B709},  307  (2005);
A. Bashir and A. Raya, Few-Body Syst. {\bf 41}, 185 (2007); A. Bashir and A. Raya, in {\em Trends in Boson Research},
edited by A.V. Ling, 1st. edition (Nova Science Publishers, Inc. N.Y., 2006), pp. 183-229. ISBN 1-59454-521-9, hep-ph/0411310;
A. Bashir, A Raya, I. Cl\"oet and C.D. Roberts, arXiv:0806.3305 [hep-ph].
%
\bibitem{graphene} V.P. Gusynin and S.G. Sharapov, Phys. Rev. Lett. {\bf 95},  146801 (2005); 
K.S. Novoselov {\em et. al.}, Nature {\bf 438},  197 (2005); 
Y. Zhang {\em  et. al}, Nature {\bf 438},  201  (2005).
%
%
\bibitem{kubo} R. Kubo, J. Phys. Soc. Jpn. {\bf 12}, 570  (1957).
%
\bibitem{Ishi} K. Ishikawa and T. Matsuyama, Nucl. Phys. {\bf B280},  523  (1987).
%
\bibitem{WGTI} J.C. Ward, Phys. Rev. {\bf 78},  182 (1950); 
H.S. Green, Proc. Phys. Soc. (London) {\bf A66},  873  (1953); 
Y. Takahashi, Nuovo Cimento {\bf 6},  371  (1957).
%
\bibitem{polyac} R. Jackiw and C. Rebbi, Phys. Rev. {\bf D13},  3398  (1976); W.P. Su, J.R. Scrieffer and A.J. Heeger, Phys. Rev. Lett. {\bf 42},  1698 (1979); R. Jackiw and J.R. Scrieffer, Nucl. Phys. {\bf B190},  253(1981).
%
\bibitem{Jackiw} R. Jackiw, Phys. Rev. {\bf D29},  2375  (1984).
%
\bibitem{Hald} F.D.M. Haldane, Phys. Rev. Lett. {\bf 61},  2015 (1988).

\bibitem{Swamy2} R. Acharya and P. Narayana Swamy, Int. Jour. Mod. Phys. {\bf A6},  861  (1994).
%
\bibitem{kurash} P.A. Kurashvili, cond-mat/0504357.
%
\bibitem{jellal} A. Jellal, Int. Jour. Theo. Phys. {\bf 37},  2187  (1998); \emph{ibid} Int. Jour. Theo. Phys. {\bf 37}, 2751  (1998).
%
\bibitem{leitner} M. Leitner, Adv. Theor. Math. Phys. {\bf 12} 479 (2008).
%
\bibitem{Chuy} Ma. de J. Anguiano and A. Bashir, Few Body Systems {\bf 37},   71 (2005).
%
\bibitem{Shimizu} K. Shimizu, Prog. Theor. Phys. {\bf 74},  610  (1985)
%
\bibitem{ABR} M. de J. Anguiano, A. Bashir and A. Raya, Phys. Rev. {\bf D76} 127702 (2007).
%
\bibitem{hatsugai} Y. Hatsugai, M. Kohmoto and Y.-S. Wu, Phys. Rev. {\bf B54},  4898 (1996).
%
\bibitem{JT} R. Jackiw and S. Templeton,  Phys. Rev. {\bf D23}, 2291 (1981).
%
\bibitem{Sharapov} S.G. Sharapov, V.P. Gusynin and H. Beck, Phys. Rev. {\bf B69}, 075104 (2004).
%
\bibitem{ddw} A.A. Nersesyan and G.E. Vachanadze, J. Low Temp Phys. {\bf 77}, 293 (1989): X. Yang and C. Nayak, Phys. Rev. {\bf B65}, 064523 (2002).
%
\bibitem{lg} G.W. Semenoff, Phys. Rev. Lett. {\bf 53}, 2449 (1984); 
J, Gonz\'alez, F. Guinea and M.A.H. Vozmediano, Nucl. Phys. {\bf B406}, 771 (1993); J, Gonz\'alez, F. Guinea and M.A.H. Vozmediano, Phys. Rev. {\bf 63}, 134421 (2001).
%
\bibitem{Kondo} K.-I. Kondo Int. J. Mod. Phys. {\bf A11},  777  (1996).
%
\bibitem{Avron} J. A. Avron and R. Seiler, Phys. Rev. Lett. {\bf 54},  259 (1985); J. A. Avron, R. Seiler and B. Shapiro, Nucl. Phys. {\bf B265},  364 (1986).
%
\bibitem{CH} S. Coleman and B. Hill, Phys. Lett. {\bf B159},  184  (1985).
%
\bibitem{khare} A. Khare, \emph{Fractional Statistics and Quantum Theory} 2nd. Edition(World Scientific, 2005), ISBN 981-256-160-9.
%
\bibitem{Del} R. Delbourgo and A. Waites, Aust. J. Phys. {\bf 47},  465 (1994).
%
\bibitem{frohlich} J. Fr\"ohlich and T. Kerler, Nucl. Phys. {\bf B354},  361 (1991).
%
\bibitem{Eisenstein} H.W. Jiang, H.L. Stormer, D.C. Tsui, L.N. Pfeiffer and K.W. West, Phys. Rev. B {\bf 40}, 12013 (1989); J.P. Eisenstein, G.S. Boebinger, L.N. Pfeiffer, K.W. West and S. He, Phys. Rev. Lett. {\bf 68},  1383 (1992); Y.W. Suen, L.W. Engel, M.B. Santos, M. Shayegan and D.C. Tsui, Phys. Rev. Lett. {\bf 68},  1379 (1992); Y.W. Suen, H.C. Manoharan, X. Ying, M.B. Santos and M. Shayegan, Phys. Rev. Lett. {\bf 72},  3405 (1994).

\bibitem{Willet} R.L. Willet, R.R. Ruel, K.W. West and L.N. Pfeiffer, Phys. Rev. Lett. {\bf 71},  3846 (1993).
%
\bibitem{HLR} B. Halperin, P.A. Lee and N. Read, Phys. Rev. {\bf B47},  7312 (1993).


\bibitem{stormer} H.L. Stormer, D.C. Tsui and A.C. Gossard, Rev. Mod. Phys. {\bf 71},  S298 (1999).
%
%
%
%
%
%
\bibitem{Lud} A.W.W. Ludwig, M.P.A. Fisher, R. Shankar and G. Grinstein, Phys. Rev. {\bf B50} 7526 (1994).
%
%
%
%
%
\bibitem{kat} M.I. Katnelson, Eur. Phys. J. {\bf B51} 157 (2006).
%
\bibitem{GusShar} V.P. Gusynin and S.G. Sharapov, Phys. Rev. {\bf B73},  245411 (2006). We thank V. Gusynin for bringing to our attention Refs.~\cite{Hald,GusShar}.

\bibitem{GusRev} V.P. Gusynin, S.G. Sharapov and J.P. Carbotte, Int. J. Mod. Phys. {\bf B21}, 4611 (2007); I.F. Herbut, Phys. Rev. {\bf B76} 085432 (2007).

\end{thebibliography}
\end{document}